\documentclass[12pt,a4paper]{article}
\usepackage{amsmath}
\usepackage{amssymb}
\usepackage{caption}
\usepackage{hyperref}
\usepackage[numbers]{natbib}
\renewcommand{\cite}[1]{\textsuperscript{\citenum{#1}}}
\begin{document}

\section*{Title} Selection of single cell clustering methodologies through rank aggregation of multiple performance measures

\section*{Authors} 
Owen Visser\footnote{University of Florida, Department of Biostatistics}, 
Somnath Datta\textsuperscript{1}

\section*{Abstract}

As single-cell gene expression data analysis continues to grow, the need for reliable clustering methods has become increasingly important.
The prevalence of heuristic means for method choice could lead to inaccurate reports if comprehensive evaluation of the methods is omitted. 
Typical comparisons of methods fail to address the complexity presented by the data, transformations, or internal parameters.
Previous work in the field of microarray data provided measures to evaluate the stability characteristic of clustering algorithms.
Additional work on aggregation in the same era presented a way to compare multiple methodologies using several performance measures. 
In this paper, we provide adaptations to these measures and employ two aggregation schemes to create ranked lists of method and parameter choices for six unique datasets.
Our findings demonstrate that an ensemble of validation measures, combined with ranking based on measures' dataset specific preferences, provides an objective way to select clustering methodologies, taking into account characteristic evaluation from each measure.

\section*{Introduction}

One aim within single-cell RNA sequencing (scRNA-seq) analysis is the identification of cell populations, commonly achieved through clustering based on gene expression data \cite{kiselev2019challenges}.
The continued growth of the scRNA-seq research field has led to many researchers developing unique pipelines, each with their own distinct clustering algorithm and pre-clustering transformations \cite{vieth2019pipelines}.
These transformations, particularly dimension reduction, influence the clustering process and the characteristics of the resulting clusters \cite{ronan2016avoiding}. 
There are no stringent guidelines in measuring these characteristics, and selection of the methodological choices is often left to the researcher to choose.
Without proper selection, one may under- or over-cluster, resulting in a lack of detection or the false identification of groups, respectively \cite{grabski2023significance}. 
Even with rigorous comparison of parameter choices, the inherent, unknown structure of scRNA-seq data prevents consistent performance of any validation measure \cite{handl2005computational, kharchenko2021triumphs}.

Typical comparisons of clustering methodologies use one or more validation measures on simulated or benchmark data. 
However, these approaches are often limited in scope. 
For example, in Emmons \cite{emmons2016analysis} comparison of four graphical clustering methods was restricted to simulated data, overlooking nuances present in real datasets. 
In another study, Patterson-Cross \cite{patterson2021selecting} used the silhouette score \cite{rousseeuw1987silhouettes} to evaluate two clustering methods on three real datasets, but the use of a single measure does not fully capture all characteristics of clustering performance.
As a final example, Qi \cite{qi2020clustering} combined the adjusted Rand index \cite{hubert1985comparing} and true positive accuracy to compare eleven single-cell analysis tools, four classification methods, and three clustering methods across twelve datasets. 
Although extensive, this study did not account for all relevant parameters choices within pre-clustering and clustering methods.

Historically, the field of microarray analysis encountered similar challenges, which were addressed through several stability measures \cite{datta2003comparisons} and a cross-entropy approach to rank aggregation \cite{pihur2007weighted}.
Pihur\cite{pihur2007weighted} showed that the combination of stability measures with traditional clustering validation measures allowed for a more complete evaluation of the characteristics of the clusters in microarray analysis studies.
Unfortunately, this evaluation technique is not directly applicable to scRNA-seq data.
The measures developed by Datta\cite{datta2003comparisons} involve perturbing microarray data by removing samples of bulk cells, leading to a partial loss of information in the genes being clustered.
To follow the equations precisely for cell clustering, genes should be removed to evaluate the resulting clusters.
However, completely ignoring the expression of a gene within cells would not achieve the same effect; particularly in the presence of feature selection or dimension reduction.

In this paper, we define novel adaptations of stability measures from the microarray field through a k-fold cross-validation approach \cite{schaffer1993selecting}, ensuring the retention of original functionality.
We pair these adapted measures with several traditional validation measures to evaluate clustering methodologies at varying parameter values.
Additionally, we construct two aggregation schemes using \verb|RankAggreg| to compare all choices simultaneously, or within cluster size, to identify the ten optimal method choices across six datasets \cite{pihur2009rankaggreg}.
Our work provides an objective and decisive approach for choosing clustering methods and parameters based on performance measures in scRNA-seq analysis.

\section*{Results}\label{results}

\subsection*{Gold Standard Data}
We first consider five real gold standard datasets, where gold-standard indicates that rigorous analysis of all cells has been conducted for the entirety of each dataset and true cell identity is largely agreed upon \cite{qi2020clustering}. 
The Biase \cite{biase2014cell} dataset contains three groups of forty-eight mouse embryo cells. 
The Deng \cite{deng2014single} dataset contains ten distinctions of 268 mouse pre-implantation embryos and mature cells. 
The Goolam \cite{goolam2016heterogeneity} dataset contains five groups of 124 mouse embryo cells. 
The Pollen \cite{pollen2014low} dataset contains four broad groups and eleven sub-groups of 301 cells derived from developing human cortex. 
The final dataset, Yan \cite{yan2013single}, contains eight groups of 124 cells from human pre-implantation embryos and human embryonic stem cells. 
Initially, each of the five datasets were pre-processed as described in the Methods section, clustered as a full set, and each partition was clustered such that for each partition a single, unique cell was removed once.
Each method was given an abbreviation as follows:
Principle Component Analysis (PCA) \cite{pearson1901liii} and Hierarchical (agglomerative) using the Wards Linkage (HA) \cite{ward1963hierarchical} is denoted as PHC, PCA and K-means \cite{hartigan1979algorithm} as PKM, PCA and Louvain\cite{blondel2008fast} as PLV, PCA and Leiden\cite{traag2019leiden} as PLN, t-distributed Stochastic Neighbor Embedding (tSNE) \cite{van2008tsne} and HA as THC, tSNE and K-means as TKM, tSNE and Louvain as TLV, tSNE and Leiden as TLN, RaceID\cite{grun2015single} as RID, pcaReduce\cite{vzurauskiene2016pcareduce} using statistical merging as PRS, pcaReduce using sampling merging as PRM, and CIDR\cite{lin2017cidr} remains the same.

Each method was evaluated at a set of parameters for each dataset. 
The parameter of cluster size ($cs$) was evaluated in the ranges $cs \in \{2,\dots,7\}$, $cs \in \{7,\dots,13\}$, $cs \in \{2,\dots,8\}$, $cs \in \{2,\dots,14\}$ , and $cs \in \{5,\dots,12\}$ for the Biase, Deng, Goolam, Pollen, and Yan datasets, respectively.
As for the graphical clustering methods, Leiden and Louvain, the number of nearest neighbors ($nn$) was evaluated at $nn \in \{10, 25, 40\}$, and resolution ($res$) was evaluated at each $0.1$ step on the range of $(0,1)$.
We define methodologies as the unique pairing of method and parameter values, and will henceforth refer to methods as methodologies, and vice versa, interchangeably.
The inclusion of Leiden and Louvain methodologies were restricted such that only parameter choices which created clusters within the ranges of the cluster size parameter were retained. 
This resulted in 132, 66, 93, 139, and 102 methodologies for the Biase, Deng, Goolam, Pollen, and Yan datasets, respectively.

When considering the measures individually, it becomes clear that no measure can be relied upon alone to give consistent results, as seen in Tables \ref{clustersizes_all} and \ref{clustersizes}.
Each measure tends to favor either high or low cluster sizes for the datasets considered, some even swapping between highest and lowest (e.g., ADM and DI).
Additionally, none of the measures could consistently place a method with optimal cluster sizes---equal to the number of true groups--- in the top five list; even for external measures. 
Notably, the Average Distance (AD) did not identify any methods with optimal cluster sizes, and the common measures of Silhouette Width (SW) and Dunn Index (DI) could only identify methods for one dataset.

Employing the fast rank aggregation scheme, A, for the Pollen dataset, the top 5 ranked methods are PRS (cs=3), PRS (cs=5), PRM (cs=5), PRM (cs=4), and PRM (cs=3), as seen in Table \ref{onestep}.
These cluster sizes closely resemble the number of known broad groups (4).
Methods ranked by scheme A for the Biase dataset showed preference to smaller cluster sizes as well, for the first four ranks the lowest cluster size is the highest with the top ranked method being RID (cs=2). The fifth ranked method is PHC (cs=3), with complete congruence compared to the true groups.
Similar rankings follow with other datasets, where the top ranked method are lower in cluster size, but with the true number of groups being within the top 5 methods chosen.
The ranked lists produced by scheme A for the Deng and Yan datasets contain methodologies whose cluster sizes match the true number of biological groups with TLV (nn=25, res=0.2) and RID (cs=8) both at rank 3, respectively.
The Goolam dataset did not contain a method with a cluster size equal to that of the true groups. 
Instead, smaller cluster sizes showed at the top ranks, with method such as RID (cs=3).

In scheme B, each dataset included a cluster size which matched its true number of groups in its top five list.
For the Pollen dataset, the top rank is now PKM (cs=10), close to the correct amount but with the best performing ARI of all methods considered.
In addition, PRS (cs=11) is comparable to the true number of groups at rank 3, which is higher ranked than any individual measure. 
Interestingly, the cluster sizes of the top methods in Table \ref{twostep} more closely compare to the number of subgroups, rather than broad groups from the first scheme.
The scheme B ranked lists for the Biase dataset held CIDR (cs=3) at rank 1; correctly identifying all cells.
The ranked list for the Goolam dataset held the optimal cluster size as the top ranked method, RID (cs=5).
Deng's list contains the optimal number of clusters (10) with TLV (nn=25, res=0.2) at rank 3.
For the Yan dataset list, the second ranked method was the optimal choice RID (cs=8), whereas the top ranked method was RID (cs=7).

\subsection*{Mouse Bone Marrow Data}
A dataset of mouse bone marrow cells was used to demonstrate how to utilize the schemes in methodological choice when biological truth is unknown. 
The dataset consists of 2256 endosteal bone marrow cells from mice at the age of 3months \cite{yao2020_3m}.
The bone marrow dataset was initially cleaned by removing cells whose gene expression showed more than 5\% of genes mapping to mitochondrial genome (determined by the R package \verb|org.Mm.eg.db| \cite{org.Mm.eg.db}), cells with less than 200 gene count, and cells with more than 6000 gene count. 
Next, the genes were filtered such the top 2000 most variable genes were retained.
Following this, using the R package \verb|scuttle| \cite{scater}, the 2256 remaining cells had their gene counts log-normalized and a pseudo count was added.
After cleaning, 100 subsets of the data were made from the removal of equal sized partitions of cells which were randomly chosen without replacement.
Each subset was clustered using the same clustering methods and abbreviations as the gold-standard datasets. 
The parameter choices for the clustering methods are as follows: The parameter cluster size was evaluated at $cs \in \{2, \dots, 30\}$, the number of nearest neighbors was evaluated at $nn \in \{20, 30, 40, 50\}$, and resolution was evaluated at each $0.1$ step on the range of $(0,1)$.
The inclusion of Leiden and Louvain methodologies were restricted such that only parameter choices which created clusters within the ranges of the cluster size parameter were retained. 
This resulted in 291 methodologies whose cluster sizes were within the range of [2, 30].
Tables \ref{onestep mouse} \& \ref{twostep mouse} provide the resulting top 5 ranked lists of methodologies for scheme A and B, respectively. 
It can be seen that cluster size 2 made up the top list for scheme A.
Alternatively, the unique clusters sizes presented by scheme B show that the best methods occur at clusters sizes 2, 3, 10, 11, and 5, in order. 
The broad grouping of THC (cs=2) finds itself in the top spot in both schemas ranked lists. 
There may be some overlying structure to the data that is being captured by the measures for these large clusters.
In scheme B, methods like RID (cs=10) and RID (cs=11) offer a more refined view of the cluster structures.



\section*{Discussion}\label{discuss}

As presented in this paper, no single clustering method is universally optimal, nor can a single measure supplement the task of identifying the overall best performing clustering method for multiple datasets.
The measures needed to identify the best method, and the method itself, will vary depending on the specific characteristics of a dataset.
Aggregation of a diverse set of validation measures provides an objective approach to method choice and more thorough evaluation of clustering methodologies compared to a single validation measure.  
In each case considered, scheme B was able to place a methodology whose cluster size matched the true number of groups while individual measures could not.

Undoubtedly, there is more work to be done in the field of scRNA-seq data cluster validation. 
For example, feature selection in single-cell data analysis was not explicitly addressed for any dataset. 
Different feature selection techniques, such as highly variable gene selection or mutual information-based methods, have a significant impact on the clustering results \cite{vieth2019pipelines}. 
It is commonplace among cluster validation research in this field to omit feature selection techniques \cite{kiselev2019challenges, qi2020clustering, duo2018systematic, patterson2021selecting}.
While we did not approach this level of comparison, the framework presented is adaptable, with the potential to incorporate additional parameter-like selections within methods. 

The measures utilized did not cover the entire spectrum of characteristics that internal and external measures can cover; more can be seen in work by Hassan\cite{hassan2024z}. 
However, the list of measures can be easily updated to include metrics that serve to validate additional characteristics of clusters resulting from different methodologies and parameter settings.
Moreover, adjustment can be done to the schema as well with the introduction of weighting for different measures. 
Pihur recommends a uniform weighting across measures which evaluate similar characteristics \cite{pihur2007weighted}. 
Another weighting scheme could consider the tendency some measures have towards cluster size, either concordant or non-concordant, as is seen in most clustering validation measures \cite{handl2005computational}. 

The aggregation scheme can also be adapted by introducing different levels of comparison for methodology choice.
The cases we considered created a direct ranking of the best performing methods in terms of overall cluster characteristics (scheme A), and provided a ranked list of the best performing methods within each cluster size, ordered by the measures of those characteristics within and across the unique cluster sizes (scheme B).  
One could create additional schema to compare different levels of parameter choice, such as feature selection or dimension count. 

In conclusion, the ensemble of validation measures introduced in this paper, along with the two aggregation schemes, provide an objective framework for selecting clustering methods in scRNA-seq data analysis.  

\section*{Methods}\label{methods}

\subsection*{Measures}
We adapt measures presented by Datta\cite{datta2003comparisons} by utilizing a $k$-fold cross-validation \cite{schaffer1993selecting} strategy to perturb the cells being clustered.
For a single fold, consider the duplication of a scRNA-seq expression matrix.
In the first, a partition of cells is removed before dimension reduction and clustering, resulting in fewer cells in the cluster assignment.
In the second, the same partition is removed after dimension reduction and clustering, ensuring congruence between the cells indices.
Comparing the two resulting assignment vectors serves as a measure of stability when this partition is removed.
This stability measure can be extended to the average of $k$ many folds where in each a partition is removed before and after dimension reduction and clustering.
\subsubsection*{Notation}

Consider a scRNA-seq gene expression matrix with $g$ genes and $l$ cells. The set of all cells, denoted as $L$, can be separated into $N$ true groups, and to $M$ clusters through a clustering method.
Each cluster can be referenced either by the $i^\text{th}$ cell it contains (e.g., the cluster containing cell $i$ is denoted as $C(i)$, where $i = 1, \dots, l$) or by an arbitrary index of the $m^\text{th}$ cluster (e.g., the $m^\text{th}$ cluster is denoted as $C_m$, where $m = 1, \dots, M$).
Additionally, consider $K$ alternate versions of the data where, for each version, a single equal sized partition, $J_k$, is removed, where $k= 1,\dots,K$.
When a cluster is formed on data perturbed by the removal of partition $J_k$, it is denoted as $C^{J_k}$.
Conversely, when the partition $J_k$ is removed after the clustering method is applied, the resulting clusters are denoted as $C^{-J_k}$.
Using the notation above each measure has been defined.
Although all traditional measures can be directly applied to scRNA-seq data, they have been defined under our notation for the sake of completeness. 


\subsubsection*{Adjusted Rand Index}
The Adjusted Rand Index (ARI) is a measure of the agreement between two sets of cluster assignments corrected for chance \cite{hubert1985comparing}. 
The ARI is defined as follows:
\begin{align}\label{ARI}
 ARI &= \frac{
 \sum_{i,h}^l \binom{l_{i,h}}{2} - \left[ \sum_{i} \binom{a_i}{2}  \cdot\sum_{h} \binom{b_h}{2} \right] / \binom{l}{2}
 }{
 \frac{1}{2} \left[ \sum_{i} \binom{a_i}{2}  + \sum_{h} \binom{b_h}{2} \right] -  \left[ \sum_{i} \binom{a_i}{2} \cdot \sum_{h} \binom{b_h}{2} \right] / \binom{l}{2}
 },
\end{align}
where $l$ is the number of cells, $a$ and $b$ are different sets of cluster assignments, $a_i$ denotes the number of cells in cluster $i$, $b_h$ denote the number of cells in cluster $h$, and $l_{i,h}$ is defined as number of cells in cluster $i$ for set $a$ and cluster $h$ for set $b$.
The ARI is valued on the interval $[0,1]$, where $0$ indicates no agreement, and $1$ indicates complete agreement.

\subsubsection*{Average Distance}
The average distance (AD) evaluates the consistency of cell's cluster assignments across data perturbation via a distance metric (Euclidean, Manhattan, etc.). 
More specifically, AD calculates the distance between cell's gene expression profiles contained in the intersecting clusters across the perturbed and non-perturbed data \cite{datta2003comparisons}.
The AD is defined as, 
\begin{align}\label{AD}
    AD =  \frac{1}{K}\sum_{k=1}^{K}
    \frac{1}{n(L\setminus J_k)}\sum_{h\in L\setminus J_k} \bigg(&\frac{1}{n\big(C^{J_k}(h)\big) n\big(C^{-J_k}(h)\big)}\times \\ &\sum_{\substack{i \in C^{-J_k}(h)\\ i' \in C^{J_k}(h)}} d(x_i, x_{i'}) \bigg),
\end{align}
where $x_i$ is the gene expression profile of cell $i$, and $d(x_i, x_{i'})$ is the euclidean distance in the gene expression profiles of cell $i$ and cell $i'$.
The AD is valued on the interval $[0,\infty)$, where lower values indicate congruence in cluster assignments.

\subsubsection*{Average Distance Between Means}
The average distance between means (ADM) evaluates the consistency of average cell cluster's gene expression profiles over data perturbation \cite{datta2003comparisons}. 
The ADM is defined as,
\begin{gather}\label{ADM}
    ADM = \frac{1}{K} \sum_{k=1}^{K} \frac{1}{n(L\setminus J_k)}\sum_{i\in L\setminus J_k} d\left(
    \bar{x}_{C^{J_k}(i)}, \bar{x}_{C^{-J_k}(i)}
    \right),
\end{gather}
where $\bar{x}_{C^{J_k}(i)}$ is the average gene expression profile of the cells clustered within $C^{J_k}(i)$, and similarly $\bar{x}_{C^{-J_k}(i)}$ within $C^{-J_k}(i)$.
The ADM is valued on the range $[0,\infty)$, where lower values indicate consistency in cluster locations.

\subsubsection*{Average Proportion of Non-overlap}
The average proportion of non-overlap (APN) evaluates the consistency of clustering across data perturbation \cite{datta2003comparisons}.
The APN is defined as, 
\begin{gather}\label{APN}
    APN =\frac{1}{K} \sum_{k=1}^{K} \frac{1}{n(L\setminus J_k)}\sum_{i\in L\setminus J_k} \left( 1 - \frac{n\big(C^{-J_k}(i) \cap C^{J_k}(i)\big) }{ n\big(C^{-J_k}(i)\big)} \right),
\end{gather}
where $n(A)$ is the cardinality of a set $A$, and $A\cap B$ is the intersection of sets $A$ and $B$.
APN is valued on the interval $[0,1]$, where lower values indicate more consistent clustering of cells.


\subsubsection*{Biological Homogeneity Index}
The Biological Homogeneity index (BHI) measures the homogeneous nature of statistical clusters of cells and their true biological grouping. 
Let $T(x)$ and $T(y)$ be known cell groups containing cell $x$ and $y$, respectively.
The BHI is then defined as,
\begin{align}\label{BHI}
    BHI = \frac{1}{M} \sum_{m=1}^{M} \frac{1}{n(C_m)(n(C_m) - 1)} \sum_{x\neq y \in C_{m}} I\big(T(x) = T(y)\big),
\end{align}
where the indicator function $I(T(x) = T(y))$ equals 1 when $T(x)$ and $T(y)$ are equivalent, and zero otherwise.
The BHI is valued on the interval of $[0,1]$ and indicates more homogeneous clusters as the measure approaches 1 \cite{pihur2009cluster}.

\subsubsection*{Biological Stability Index}
The Biological Stability Index (BSI) measures the consistency in assigning similar biological groups to cells \cite{datta2006methods}.
More specifically, BSI calculates the average proportion of overlap given by the perturbation of data within a known biological group.
The BSI is defined as,
\begin{align}\label{BSI}
    BSI =  \frac{1}{K}\sum_{k=1}^{K} \bigg[ 
    \frac{1}{N^{J_k}}\sum_{i=1}^{N^{J_k}}\bigg( &
    \frac{1}{\big(n\big(T^{J_k}_i\big) - 1\big)n\big(T^{J_k}_i\big) } \times\\ &\sum_{x\neq y \in T_i^{J_k}} \frac{n\big( C^{-J_k}(x) \cap C^{J_k}(y)\big)}{n\big( C^{-J_k}(x)\big)} \bigg) \bigg],
\end{align}
where $T_i^{J_k}$ is the $i^{th}$ biological group excluding partition $J_k$, and $N^{J_k}$ is the number of biological groups if partition $J_k$ is excluded.
The BSI is valued on the interval of $[0,1]$, where values closer to 1 indicate more stable clustering in regards to the biological grouping of cells \cite{datta2006methods}.

\subsubsection*{Connectivity}
Connectivity (CN) measures the ranked distance of cells if they are not assigned the same cluster \cite{pihur2009cluster}.
Connectivity is defined as, 
\begin{align}\label{CN}
   CN = \sum_{i=1}^l \sum_{h=1}^{H} \delta_{i, nn_{i(h)}},
\end{align}
where $H$ is a user-defined integer with restriction $0<H<l$, $nn_{i(h)}$ is defined as the $h^\text{th}$ nearest neighbor of cell $i$, and $\delta_{i, nn_{i(h)}}$ is an indicator that takes the value zero when $C(h)$ is equal to $C(i)$, and $1/h$ otherwise.
Connectivity is valued on the interval [0, $\infty$) where smaller values indicate better cell grouping based on gene expression profile similarity.

\subsubsection*{Dunn Index}
The Dunn index (DI) measures the ratio of the smallest distance between gene expression profiles of cells not in the same cluster to the largest intra-cluster distance \cite{dunn1974well, pihur2007weighted}.
The Dunn index is defined as
\begin{align}\label{DI}
    DI = \dfrac{\min\limits_{C_i, C_h \in \mathcal{C}; C_i \neq C_h} \left( \min\limits_{s\in C_i, t\in C_h} d(x_s, x_t) \right)}{ \max\limits_{C_r\in\mathcal{C}} \left( \max\limits_{s\neq t \in C_r} d(x_s, x_t) \right)},
\end{align}
where $\mathcal{C} = \{C_1, \dots, C_M\}$ is the set of all clusters.
The DunnIndex is valued on the interval [0, $\infty$) and should be maximized.

\subsubsection*{In-group Proportion}
The in-group proportion (IGP) measures the proportion of cells whose nearest neighbor are assigned to the same cluster \cite{kapp2007clusters}. 
We choose to follow the design of \cite{pihur2009cluster} and calculate the overall IGP score, which is defined as 
\begin{align}\label{IGP}
IGP = \sum_{m=1}^M \frac{n(i :i \in C_m \text{ and } i^* \in C_m )}{n(C_m)},
\end{align}
where $i^*$ is the nearest neighbor of $i$ defined as $i^* = argmin_{h;h\neq i}d(x_i, x_h)$.
The overall IGP score is valued on the interval $(0,M)$ and larger scores correspond to better predictive ability \cite{pihur2009cluster}.

\subsubsection*{Silhouette Width}
Silhouette width (SW) \cite{rousseeuw1987silhouettes} is the average of each cells silhouette value.
The Silhouette value of cell $i$ is defined as 
$$
S(i) = \frac{b_{i} - a_{i}}{\max(a_{i}, b_{i})},
$$
where $a_i$ is the average distance between the gene expression profiles of cell $i$ and all other observations in the same cluster, and $b_i$ is the average distance between the gene expression profile of cell $i$ and the gene expression profiles of the cells in the nearest cluster by distance.
The values $a_i$ and $b_i$ are defined as
$$
a_{i} = \frac{1}{n(C(i)) - 1} \sum_{h \neq i \in C(i)} d(x_i, x_h),
$$
and
$$
b_{i} = \min_{C_k \in \mathcal{C} \backslash C(i) }  \frac{1}{n(C_k)}  \sum_{h\in C_k} d(x_i, x_h).
$$
The Silhouette width of the entirety of cell clusters can be defined as,
\begin{align}\label{SW}
    SW = \frac{1}{l}\sum_{i=1}^l S(i).
\end{align}
SW is valued on the interval $[-1,1]$, where values closer to 1 indicate better separation of clusters and more compact clusters, and values closer to -1 indicate loosely packed and overlapping clusters.

\subsection*{Clustering}\label{ClusteringMethods}

\subsubsection*{Pre-clustering}
To ensure consistency in our comparison of methods, we implemented a standard data-cleaning process: filtering genes with no counts, log-normalizing the data, and adding a pseudo count, all performed using the R package \verb|scuttle| \cite{scater}.

Among the dimension reduction techniques employed by the methods that will be compared are Principal Component Analysis (PCA) \cite{pearson1901liii} and t-SNE \cite{van2008tsne}.
PCA reduces dimensionality by transforming the original data into a new set of orthogonal variables (principal components), ordered by the level of variance they explain from the original data.
t-SNE reduces dimensionality by mapping high-dimensional data points into a lower-dimensional space (typically two---sometimes three---dimensions) in a way that preserves local structures and patterns; primarily pairwise point distances.
Each of these techniques were used in conjunction with traditional clustering methods and conducted in R using the packages \verb|stats| \cite{statsR} and \verb|Rtsne| \cite{van2008tsne}, respectively. 
Methods that did not use PCA and tSNE have their dimension reduction techniques discussed within the method descriptions. 

\subsubsection*{Clustering Methods}
\paragraph{K-means}
The \textit{k-means} \cite{hartigan1979algorithm} divides a dataset into $K$ distinct, non-overlapping subsets based on distance. 
The algorithm initializes by randomly selecting $K$ points as cluster centroids and then assigns each remaining data point to the nearest centroid. 
After this the algorithm will iteratively recalculate the location of the centroids as the average distance of its cluster's members and reassign the nearest cells until no changes occur. 
The parameter of interest in this method is the number of clusters.

\paragraph{Hierarchical}
The \textit{hierarchical} algorithm \cite{ward1963hierarchical} creates a tree of clusters that splits based on dissimilarity---distance in this case---either through a bottom-up approach (agglomerative), or a top-down approach (divisive).
We chose to utilize only the agglomerative method of hierarchical clustering.
Assignment to a set number of clusters was done by cutting the dendrogram at the appropriate height based on the Ward linkage \cite{ward1963hierarchical} by using \verb|cutree| in the R package \verb|stats| \cite{statsR}.
The parameter of interest in this method is the number of clusters.

\paragraph{Louvain}
The \textit{Louvain} \cite{blondel2008fast} clustering algorithm is a heuristic graphical clustering method based on optimizing modularity in communities.
The algorithm starts by assigning each node to its own community and then merges communities that increase modularity, followed by aggregating the network and repeating the process until convergence.
We first construct a k-nearest neighbor graph from the dimension-reduced cell matrix to be used as input for the Louvain clustering method using the R package \verb|cccd| \cite{cccd}.
Afterward, the R package \verb|igraph| \cite{igraph} is used to conduct Louvain clustering.
The parameters of interest within this method are the number of nearest neighbors and the resolution of the communities.

\paragraph{Leiden}
The \textit{Leiden} \cite{traag2019leiden} algorithm follows a similar process to the Louvain algorithm but adds a refinement step to ensure that communities are well-connected internally.
The construction of the k-nearest neighbor graph, implementation of the algorithm, and parameters of interest remain the same as the Louvain algorithm.

\paragraph{pcaReduce}

\textit{pcaReduce} \cite{vzurauskiene2016pcareduce} is a hierarchical (agglomerative) clustering method that iteratively clusters cells using the k-means algorithm, followed by merging clusters through a hierarchical algorithm.
The merging is guided by two criteria: the highest probability of cluster configuration either before or after merging, based on distances in a dimension-reduced space and assuming the clusters follow a multivariate Gaussian distribution.
The parameter of interest in this method is the number of clusters.

\paragraph{RaceID}

The \textit{RaceID} \cite{grun2015single} clustering method reduces data dimensionality by constructing a distance matrix, where each entry is calculated as one minus the Pearson correlation coefficient between pairs of cells.
The k-means algorithm is then applied to this distance matrix using the Euclidean metric. 
Cluster assignments are finalized by assessing their reproducibility through bootstrapping.
The parameter of interest in this method is the number of clusters.

\paragraph{CIDR}
\textit{CIDR} \cite{lin2017cidr} is an extremely efficient clustering method that reduces the dimension of the data twice; first through the removal of genes determined by a kernel density estimation of the expression counts, then through principle coordinates analysis \cite{anderson2003canonical}.
The lower-dimensional data is then clustered using the agglomerative hierarchical clustering method with Ward linkage.
The parameter of interest in this method is the number of clusters.

\subsection*{Aggregation Schemes}
The number of measures produced for each clustering methodology is extensive, making visual comparison impractical. 
To facilitate comparison, the R package \verb|RankAggreg| was utilized to rank validation measure values across all methods for each dataset.
Two schemes were developed to provide flexibility in methodology selection past the top ranked method. 
The first scheme (A) is a direct comparison of all methods measure values, producing a ranked list that may include methods with the same cluster sizes. 
The second scheme (B) compares only the best methodologies at each cluster size, producing a ranked list of methods with unique cluster sizes.
Specifically, scheme B aggregates the rankings of each method through \verb|RankAggreg| at each cluster size based on the measure values.
Afterward, the only top scoring methodology at each cluster size is aggregated by its measures.
This scheme offers a user an ordered list of the best methodologies at unique cluster sizes at the cost of computational efficiency.

When finding the optimal list while restricting input to be a specific cluster size, the descending rank that methods occupy differs.
This causes scheme B to result in slightly different ordering of best methods when compared to scheme A for methods sharing cluster sizes. 
Moreover, the cross-entropy algorithm within \verb|RankAggreg| may fail to find a unique solution when comparing multiple short lists, for instance in similar cluster size comparison or comparison across limited numbers of cluster sizes, which allows for some variation in the ranked terms (for more detail, see \cite{pihur2007weighted}).
For reproducibility, \verb|RankAggreg| was parameterized to produce lists of the top 20 ranked methodologies across all measures in the scheme A and the first part of scheme B; the second part of scheme B is limited to the number of clusters compared.
From the final ranked lists of each scheme, only the top 5 methodologies are retained, resulting in consistent lists of ranked values on repeated function uses.

\section*{Data Availability}
The raw datasets used in this paper were sourced from and are available in the original authors works. For direct access to the sequencing data please refer to the following links and codes. The raw sequencing data from the Biase\cite{biase2014cell} dataset can be found on the NCBI Gene Expression Omnibus (GEO; http://www.ncbi.nlm.nih.
gov/geo/) under accession number GSE57249. The Deng\cite{deng2014single} dataset can be found in the GEO database under accession number GSE45719 and the Sequence Read Archive under accession number SRP020490. The Goolam\cite{goolam2016heterogeneity} dataset can be found under in the Functional Genomics Data Collection under the accession number ArrayExpress: E-MTAB-3321. The Pollen\cite{pollen2014low} dataset can be found in the NCBI Sequence Read Archive (http://www.ncbi.nlm.nih.gov/
Traces/sra/) under accession number SRP041736. The Yan\cite{yan2013single} dataset can be found in the GEO database under accession number GSE36552. The mouse bone marrow data can be found in the GEO database under accession code GSE145477 and the Broad Institute Cell Portal under study number SCP1017.

\section*{Code Availability}

All code used in this paper is available publicly on at \url{https://doi.org/10.5281/zenodo.13983905}

\bibliography{reference} 

\section*{Acknowledgements}

Will probably not include.

\section*{Author Contributions}
S.D. conceived of the presented idea and supervised the findings of this work. O.V. developed the theory, performed the computations, and verified the analytical methods.

\section*{Competing Interests}
The authors declare no competing interests.

\section*{Tables}

\begin{table}[ht]
\caption{Top 5 selection of true groups}
\footnotesize
\centering
\begin{tabular}{p{.15 \textwidth} p{.04\textwidth} p{.04\textwidth} p{.04\textwidth} p{.04\textwidth} p{.04\textwidth} p{.04\textwidth} p{.04\textwidth} p{.04\textwidth} p{.04\textwidth} p{.04\textwidth}}
  \hline
 Dataset & AD & ADM & APN & CN & SW & DI & IGP & ARI & BHI & BSI \\ 
  \hline
Biase & $\Box$ & $\boxtimes$ & $\boxtimes$ & $\Box$ & $\Box$  & $\Box$ & $\boxtimes$ &$\boxtimes$ & $\boxtimes$ & $\boxtimes$\\
Deng & $\Box$ & $\boxtimes$ & $\boxtimes$ & $\boxtimes$ & $\Box$  & $\Box$ & $\boxtimes$ &$\Box$ & $\Box$ & $\boxtimes$\\
Goolam & $\Box$ & $\Box$ & $\Box$ & $\Box$ & $\Box$  & $\Box$ & $\Box$ &$\boxtimes$ & $\Box$ & $\Box$\\
Pollen & $\Box$ & $\Box$ & $\Box$ & $\Box$ & $\Box$  & $\Box$ & $\Box$ &$\boxtimes$ & $\boxtimes$ & $\Box$\\
Yan & $\Box$ & $\Box$ & $\Box$ & $\Box$ & $\boxtimes$  & $\boxtimes$ & $\Box$ &$\Box$ & $\Box$ & $\Box$\\
   \hline
\end{tabular}
\caption*{\footnotesize Illustrating the selection of methodologies whose cluster sizes match the true number of groups in the top five ranked list for individual measures. $\Box$ indicates no selection, and $\boxtimes$ indicates the selection of at least one method in the top five.}
\label{clustersizes}
\end{table}

\begin{table}[ht]
\caption{Best performing methods cluster sizes by measures}
\centering
\footnotesize
\begin{tabular}{p{.1 \textwidth} p{.04\textwidth} p{.04\textwidth} p{.04\textwidth} p{.04\textwidth} p{.04\textwidth} p{.04\textwidth} p{.04\textwidth} p{.04\textwidth} p{.04\textwidth} p{.04\textwidth}}
  \hline
Dataset & AD & ADM & APN & CN & SW & DI & IGP & ARI & BHI & BSI \\ 
  \hline
Biase & 7 & 2 & 2 & 2 & 2 & 2 & 2 & 3 & 3 & 2 \\ 
Deng & 13 & 13 & 10 & 8 & 8 & 13 & 8 & 7 & 13 & 8 \\ 
Goolam & 8 & 2 & 2 & 2 & 2 & 7 & 2 & 5 & 8 & 2 \\
Pollen & 14 & 2 & 2 & 2 & 10 & 9 & 4 & 10 & 13 & 2 \\ 
Yan & 12 & 9 & 5 & 5 & 8 & 8 & 5 & 7 & 12 & 7 \\ 
   \hline
\end{tabular}
\caption*{\footnotesize Cluster sizes of the best performing methodologies for each individual measure value for all dataset.}
\label{clustersizes_all}
\end{table}

\begin{table}[ht]
\caption{Top 5 methods for Pollen dataset under scheme A}
\footnotesize
\centering
\begin{tabular}{p{.0075\textwidth} p{.04\textwidth} p{.02\textwidth} p{.04\textwidth} p{.04\textwidth} p{.04\textwidth} p{.04\textwidth} p{.04\textwidth} p{.04\textwidth} p{.04\textwidth} p{.04\textwidth} p{.04\textwidth} p{.04\textwidth}}
  \hline
 R & M & cs & AD & ADM & APN & CN & SW & DI & IGP & ARI & BHI & BSI \\ 
 \hline
1 & PRS  & 3 & 0.07 & 1.00 & 1.00 & 1.00 & 0.62 & 0.48 & 0.99 & 0.26 & 0.39 & 1.00 \\ 
  2 & PRS  & 5 & 0.19 & 1.00 & 1.00 & 1.00 & 0.69 & 0.62 & 0.99 & 0.55 & 0.70 & 1.00 \\ 
  3 & PRM  & 5 & 0.19 & 1.00 & 1.00 & 1.00 & 0.69 & 0.62 & 0.99 & 0.55 & 0.70 & 1.00 \\ 
  4 & PRM  & 4 & 0.14 & 1.00 & 1.00 & 1.00 & 0.67 & 0.62 & 0.99 & 0.37 & 0.57 & 1.00 \\ 
  5 & PRM  & 3 & 0.07 & 1.00 & 1.00 & 1.00 & 0.62 & 0.48 & 0.99 & 0.26 & 0.39 & 1.00 \\ 
   \hline
\end{tabular}
\caption*{\small Top 5 list of measure values for methods from scheme A for the Pollen dataset, where 'R' is rank, 'M' refers to the methodology, and 'cs' is cluster size.}
\label{onestep}
\end{table}

\begin{table}[ht]
\caption{Top 5 methods for Pollen dataset under scheme B}
\footnotesize
\centering
\begin{tabular}{p{.0075\textwidth} p{.04\textwidth} p{.02\textwidth} p{.03\textwidth} p{.02\textwidth} p{.04\textwidth} p{.04\textwidth} p{.04\textwidth} p{.04\textwidth} p{.04\textwidth} p{.04\textwidth} p{.04\textwidth} p{.04\textwidth} p{.04\textwidth} p{.04\textwidth}}
  \hline
 R & M & nn & res & cs & AD & ADM & APN & CN & SW & DI & IGP & ARI & BHI & BSI \\ 
  \hline
1 & PKM &  &  & 10 & 0.33 & 0.87 & 0.87 & 0.94 & 0.80 & 0.87 & 0.93 & 1.00 & 0.98 & 0.90 \\ 
  2 & PRS &  &  & 9 & 0.32 & 0.96 & 0.96 & 0.96 & 0.80 & 0.74 & 0.96 & 0.94 & 0.91 & 0.92 \\ 
  3 & PRS &  & & 11 & 0.35 & 0.92 & 0.91 & 0.86 & 0.77 & 0.81 & 0.87 & 0.90 & 0.98 & 0.88 \\ 
  4 & TLV & 25 & 0.2 & 8 & 0.31 & 1.00 & 1.00 & 0.98 & 0.78 & 0.68 & 0.98 & 0.86 & 0.93 & 0.99 \\ 
  5 & PRS && & 12 & 0.35 & 0.91 & 0.88 & 0.89 & 0.79 & 0.75 & 0.82 & 0.91 & 0.88 & 0.87 \\ 
   \hline
\end{tabular}
\caption*{\footnotesize Top 5 list of measure values for methods from scheme B for the Pollen dataset, where 'R' is rank, 'M' refers to the methodology, 'cs' is cluster size, 'nn' is nearest neighbors, and 'res' is resolution.}
\label{twostep}
\end{table}

\begin{table}[ht]
\caption{Top 5 methods for mouse bone marrow dataset under scheme A}
\footnotesize
\centering
\begin{tabular}{p{.0075\textwidth} p{.04\textwidth} p{.02\textwidth} p{.04\textwidth} p{.04\textwidth} p{.04\textwidth} p{.04\textwidth} p{.04\textwidth} p{.04\textwidth} p{.04\textwidth}}
  \hline
R & M & cs & AD & ADM & APN & CN & DI & IGP & SW \\ 
  \hline
1 & THC & 2 & 0.02 & 0.98 & 0.99 & 1.00 & 1.00 & 1.00 & 0.88 \\ 
  2 & PKM & 2 & 0.02 & 1.00 & 1.00 & 0.98 & 0.85 & 0.99 & 0.88 \\ 
  3 & PRM & 2 & 0.02 & 0.99 & 1.00 & 1.00 & 0.78 & 1.00 & 0.88 \\ 
  4 & PHC & 2 & 0.02 & 0.98 & 0.99 & 0.99 & 0.66 & 1.00 & 0.87 \\ 
  5 & PRS & 2 & 0.02 & 0.97 & 0.99 & 1.00 & 0.89 & 1.00 & 0.87 \\ 
   \hline
\end{tabular}
\caption*{\footnotesize Top list of measure values for methods from scheme A for the mouse bone marrow dataset, where 'R' is rank, 'M' refers to the methodology, and 'cs' is cluster size.}
\label{onestep mouse}
\end{table}

\begin{table}[ht]
\caption{Top 5 methods for mouse bone marrow dataset under scheme B}
\footnotesize
\centering
\begin{tabular}{p{.0075\textwidth} p{.04\textwidth} p{.02\textwidth} p{.04\textwidth} p{.04\textwidth} p{.04\textwidth} p{.04\textwidth} p{.04\textwidth} p{.04\textwidth} p{.04\textwidth}}
  \hline
R & M & cs & AD & ADM & APN & CN & DI & IGP & SW \\ 
  \hline
1 & THC & 2 & 0.02 & 0.98 & 0.99 & 1.00 & 1.00 & 1.00 & 0.88 \\ 
  2 & PRM & 3 & 0.06 & 0.98 & 0.99 & 0.97 & 0.89 & 0.98 & 0.75 \\ 
  3 & RID & 10 & 0.32 & 0.99 & 1.00 & 0.84 & 0.68 & 0.92 & 0.94 \\ 
  4 & RID  & 11 & 0.36 & 0.99 & 0.99 & 0.84 & 0.68 & 0.91 & 0.91 \\ 
  5 & TKM & 5 & 0.13 & 0.81 & 0.90 & 0.95 & 0.74 & 0.97 & 0.97 \\
   \hline
\end{tabular}
\caption*{\footnotesize Top list of measure values for methods from scheme B for the mouse bone marrow dataset, where 'R' is rank, 'M' refers to the methodology, and 'cs' is cluster size.}
\label{twostep mouse}
\end{table}

\end{document}